\documentclass[journal]{IEEEtran}

\usepackage[utf8]{inputenc} 
\usepackage[T1]{fontenc}
\usepackage{url, ifthen, cite}
\usepackage[cmex10]{amsmath}
\usepackage{graphicx, graphics, balance, dsfont}
\usepackage{amssymb, amsfonts, amsthm}
\usepackage{epsfig, tikz, standalone}

\newcommand{\PP}{\mathbb{P}}
\newcommand{\E}{\mathbb{E}}
\newcommand{\al}{\alpha}

\newtheorem{theorem}{Theorem}

\newtheorem{lemma}{Lemma}

\newtheorem{proposition}{Proposition}

\begin{document}
\title{Wireless Powered Mobile Edge Computing:\\ Offloading Or Local Computation?\vspace{-1mm}}

\author{Constantinos Psomas, \IEEEmembership{Senior Member, IEEE}, and Ioannis Krikidis, \IEEEmembership{Fellow, IEEE}
  
\thanks{C. Psomas and I. Krikidis are with the Department of Electrical and Computer Engineering, University of Cyprus, Nicosia, Cyprus (e-mail: \{psomas, krikidis\}@ucy.ac.cy). This work was co-funded by the European Regional Development Fund and the Republic of Cyprus through the Research and Innovation Foundation, under the projects INFRASTRUCTURES/1216/0017 (IRIDA) and POST-DOC/0916/0256 (IMPULSE). This work has also received funding from the European Research Council (ERC) under the European Union's Horizon 2020 research and innovation programme (Grant agreement No. 819819).}\vspace{-8mm}}

\maketitle

\begin{abstract}
Mobile-edge computing (MEC) and wireless power transfer are technologies that can assist in the implementation of next generation wireless networks, which will deploy a large number of computational and energy limited devices. In this letter, we consider a point-to-point MEC system, where the device harvests energy from the access point's (AP's) transmitted signal to power the offloading and/or the local computation of a task. By taking into account the non-linearities of energy harvesting, we provide analytical expressions for the probability of successful computation and for the average number of successfully computed bits. Our results show that a hybrid scheme of partial offloading and local computation is not always efficient. In particular, the decision to offload and/or compute locally, depends on the system's parameters such as the distance to the AP and the number of bits that need to be computed.
\end{abstract}\vspace{-2mm}

\vspace{-2mm}

\section{Introduction}
Emerging technologies for the development of smart homes, smart cities, intelligent transportation systems, etc., are expected to support a massive number of wireless devices (e.g., mobile phones, sensors), which will continuously exchange information. Moreover, applications such as interactive online gaming, autonomous driving and virtual reality, require real-time computational processing. This becomes a critical issue for wireless devices, given their limitations in both computational and energy resources. To overcome this constraint, mobile edge computing (MEC) \cite{MAO} and wireless power transfer (WPT) \cite{ZENG} have been proposed to support the computational and energy aspect, respectively, of such devices.

Recently, the advantages of MEC have been investigated in different communication scenarios, e.g. \cite{DING,QUEK,HUANG}. In \cite{DING}, the authors consider MEC with non-orthogonal multiple access (NOMA) for both downlink and uplink. It is shown that NOMA can reduce the latency and energy consumption of MEC offloading. The work in \cite{QUEK}, studies a scenario where a single device can offload its tasks to multiple edge servers. The proposed optimization framework minimizes the device's energy consumption as well as the execution latency of the tasks. An asynchronous MEC offloading scenario is considered in \cite{HUANG}, where the optimal resource-management policy for the task partitioning (offloading/local computation) and the time division for the transmissions is studied.

The coexistence of MEC with WPT has also been previously studied in the literature, e.g. \cite{JI,KIT,LIU,BI,WANG,ZHOU}. In \cite{JI}, an energy-efficient cooperative resource allocation policy is proposed for a wireless powered MEC system with two users; it is shown that the proposed scheme provides significant gains over systems without cooperation. MEC and WPT with cooperative communications is also studied in \cite{KIT}, with the aim of minimizing the access point's (AP's) transmit power. The work in \cite{LIU}, considers MEC and WPT in cognitive radio networks and focuses on maximizing the energy-efficiency of the devices. Energy-efficient wireless powered MEC is also  studied in \cite{WANG}, by jointly considering the energy consumption at both the energy transmitter and the user. A wireless powered multi-user MEC system is considered in \cite{BI}, where each device either computes its task locally or offloads it entirely. The authors focus on maximizing the sum computation rate for all the network's devices. Similarly, the work in \cite{ZHOU} maximizes the computation efficiency through the joint optimization of the energy harvesting time, the local computing frequency as well as the offloading time and transmit power.

In this letter, we study a point-to-point MEC system, where the device's offloading and/or local computation are wirelessly powered by an AP. In contrast to the aforementioned works, we present a mathematical framework to characterize the probability of successful computation as well as the expected number of successfully computed bits. This framework takes into account a non-linear energy harvesting model based on the diode's physics \cite{BC} and provides analytical closed-form expressions. Our results demonstrate that a hybrid partial offloading/local computation scheme is efficient but, under specific scenarios, a binary decision, i.e. full offloading or local computation, is preferable. The deterministic fading case is also considered and it is demonstrated that fading can benefit wireless powered offloading/local computation.\vspace{-1mm}

\section{System Model}\label{sys_model}
\subsubsection{Topology}
\begin{figure}[t]\centering
  \includegraphics[width=0.9\linewidth]{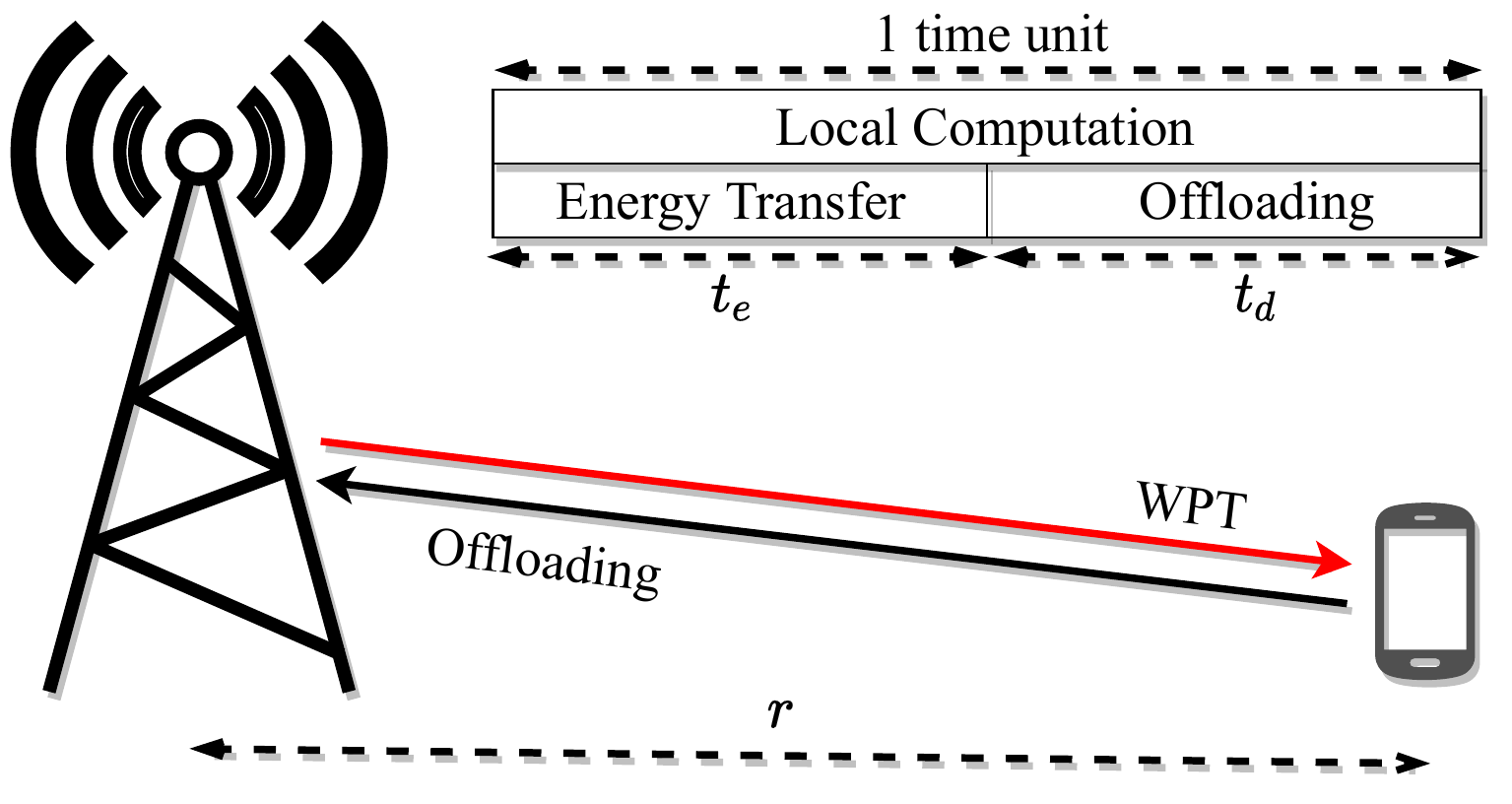}
  \caption{The considered point-to-point MEC system.}\label{model}
\end{figure}

Consider a wireless-powered communication network consisting of an AP and a device, where the AP acts as both a power beacon and a MEC server. The device is located at a distance $r$ from the AP and both are equipped with a single antenna. The AP transmits with fixed power $P$ and the device harvests energy from the AP's transmitted signal using a rectifying circuit. A harvest-then-use protocol is adopted \cite{KIT}, where all the harvested energy is used to offload the data to the MEC server and/or compute data locally. Time is slotted and a time slot duration is equal to one time unit. The network employs a time-division duplex operation and so during a time slot, the device is first in harvesting mode (downlink) for a duration $t_e < 1$ and then in offloading mode (uplink) for a duration $t_d = 1-t_e$. Note that the energy harvesting and the local computation can be done simultaneously since the energy harvester and the computing units are independent \cite{BI}. Moreover, it is assumed that the time for computation at the MEC server and for delivery of the results from the AP can be neglected \cite{KIT,ZHOU}. Fig. \ref{model} depicts the considered setup.

\subsubsection{Channel Model}
Both downlink and uplink are assumed to suffer from both small-scale block fading and large-scale path-loss effects. We consider Rayleigh block fading and so the channel coefficients are complex Gaussian distributed with zero mean and unit variance. We denote by $h \sim \mathcal{CN}(0,1)$ and $g \sim \mathcal{CN}(0,1)$, the channel coefficients for the downlink and uplink, respectively. All links exhibit additive white Gaussian noise (AWGN) with variance $\sigma^2$. The path-loss model assumes that the received power is proportional to $r^{-\al}$, where $\al > 2$ is the path-loss exponent. The AP's un-modulated transmitted signal is $s(t) = \sqrt{2 P}\Re \left\{\exp(\jmath 2\pi f_c t)\right\}$, where $\E[s^2(t)] = P$ and $f_c$ denotes the carrier frequency. Thus, the received signal at the device is
\begin{align}
y(t) &= \sqrt{2 P r^{-\al}} |h| \Re \left\{\exp(\jmath 2\pi f_c t + \jmath \theta(t))) \right\},\label{signal}
\end{align}
where $|h|$ is a Rayleigh random variable with unit parameter.

\subsubsection{Energy Transfer}
During the energy transfer phase, the device harvests radio frequency (RF) energy from the AP's transmitted signal for a duration $t_e$. The received signal is converted to direct-current (DC) using a rectifier, which is a basic circuit, usually consisting of a diode (e.g., a Schottky diode) and a passive low pass filter (LPF) \cite{ZENG}. The diode's output current from the received signal $y(t)$ can be written as $I(t) = I_s \sum_{j=1}^\infty \left(y(t)/\nu V_T\right)^j/j!$, where $I_s$ denotes the reverse saturation current of the diode, $\nu$ is an ideality factor which is a function of the operating conditions and physical contractions, and $V_T$ is the thermal voltage. By taking the expectation of $I(t)$, approximates the DC component of the current at the rectifier's output \cite{BC}. Therefore, by keeping the second and fourth order term, the total harvested energy $\epsilon$ is a non-linear function of $I(t)$, written as \cite{BC}
\begin{align}
\epsilon &= t_e\left(\gamma_2 \E[y(t)^2] + \gamma_4 \E[y(t)^4]\right)\nonumber\\
&= t_e \left(\gamma_2 P r^{-\al} |h|^2 + \frac{3}{2} \gamma_4 P^2 r^{-2\al} |h|^4\right),\label{eh}
\end{align}
where $\gamma_i$ are constants determined by the circuit's parameters $I_s$, $\nu$ and $V_T$.

\subsubsection{Task Computation}
The device decides whether to offload and/or compute locally a task of $\ell$ bits. In other words, it can either offload all of the $\ell$ bits, locally execute part of the $\ell$ bits and offload the rest (partial offloading), or compute locally the entire task.
\paragraph{Offloading} Assume that a portion of the harvested energy $\epsilon_o \leq \epsilon$, is dedicated to power the offloading to the MEC server. Then, the number of offloaded bits $\ell_o \leq \ell$ is
\begin{align}\label{lo}
\ell_o = t_d B \log\left(1+\frac{\epsilon_o}{t_d}\frac{|g|^2}{r^{\al}\sigma^2}\right),
\end{align}
where $B$ is the available bandwidth and $\epsilon_o/t_d$ is the device's transmit power. On the other hand, assuming that $\ell_o$ bits are required to be offloaded, the energy required is
\begin{align}\label{eo}
\epsilon_o = t_d \left(2^\frac{\ell_o}{t_d B}-1\right) \frac{\sigma^2 r^\al}{|g|^2}.
\end{align}
\paragraph{Local computation} Let $\psi$ be the number of cycles required by the central processing unit (CPU) to compute one-bit of data. Then, the energy consumption $\epsilon_c$ required to locally compute $\ell_c \leq \ell$ bits is \cite{HUANG}
\begin{align}\label{ec}
\epsilon_c = \xi \psi^3 \ell_c^3,
\end{align}
where $\xi$ is the effective CPU capacitance coefficient \cite{HUANG}. Hence, if $\epsilon_c \leq \epsilon$ is the energy dedicated for local computation, we have that
\begin{align}\label{lc}
\ell_c = \left(\frac{\epsilon_c}{\xi \psi^3}\right)^{\frac{1}{3}},
\end{align}
is the achieved number of locally computed bits.\vspace{-2mm}

\section{Wireless Powered MEC}\label{mec}
In this section, we derive the success probability, i.e. the probability  of successfully offloading and/or locally computing the allocated task, as well as the average number of successfully computed bits. We first provide the probability distribution of the harvested energy, which will be useful for the derivation of the aforementioned metrics.

\begin{lemma}\label{lemma}
Let $H = \theta_1|h|^2 + \theta_2|h|^4$, where $h \sim \mathcal{CN}(0,1)$ and $\theta_1, \theta_2$ are positive constants. Then, the cumulative distribution function (CDF) $F_H(x)$ of $H$ is given by\vspace{-1mm}
\begin{align}\label{cdf}
F_H(x) = 1 - \exp\left(\frac{1}{2\theta_2}\left(\theta_1 - \sqrt{\theta_1^2 + 4\theta_2x}\right)\right),
\end{align}
and the probability density function (PDF) $f_H(x)$ of $H$ is\vspace{-1mm}
\begin{align}\label{pdf}
f_H(x) = \frac{1}{\sqrt{\theta_1^2 + 4\theta_2x}} \exp\!\left(\frac{1}{2\theta_2}\!\left(\theta_1 \!-\! \sqrt{\theta_1^2 \!+\! 4\theta_2x}\right)\right),
\end{align}
respectively.
\end{lemma}

\begin{IEEEproof}
See Appendix \ref{lemma_prf}.
\end{IEEEproof}

Even though the above result refers to the case of Rayleigh fading, the extension to other fading models is straightforward by considering the corresponding probability distributions.

In what follows, we assume that the task can be split into $\ell_o = \mu \ell$ bits for offloading to the MEC server and $\ell_c = (1-\mu) \ell$ bits to compute locally, where $0 \leq \mu \leq 1$. In Theorem \ref{thm}, we derive the success probability $P_s(\mu,t_e)$, given by
\begin{align}\label{ps}
P_s(\mu,t_e) = \PP(\epsilon > \epsilon_o + \epsilon_c),
\end{align}
where $\epsilon_o$ and $\epsilon_c$ are given by \eqref{eo} and \eqref{ec}, respectively. In other words, it is the probability that the device has harvested enough energy to offload and locally compute the required bits.

\begin{theorem}\label{thm}
The probability of successfully offloading $\ell_o = \mu \ell$ bits and locally computing $\ell_c = (1-\mu) \ell$ bits, $0 \leq \mu \leq 1$, is
\begin{align}\label{sp}
  P_s(\mu,t_e) = \int_0^\infty\! &\exp\!\left(\frac{r^\al\gamma_2}{3\gamma_4 P}\left(1\!-\!\sqrt{1\!+\!\frac{6\gamma_4}{t_e\gamma_2^2}\left(\frac{\phi_o}{g}\!+\!\phi_c\right)}\right)\right)\nonumber\\
  &\times\exp(-g) dg,
\end{align}
where $\phi_o = \left(2^\frac{\ell_o}{t_d B}-1\right) t_d \sigma^2 r^\al$ and $\phi_c = \ell_c^3 \xi \psi^3$.
\end{theorem}

\begin{IEEEproof}
See Appendix \ref{thm_prf}.
\end{IEEEproof}

From Theorem \ref{thm}, we can see the trade-off between WPT and offloading/local computation. On the one hand, the device can utilize the harvested energy to offload the data to a MEC server with high computational resources rather than doing it locally. On the other hand, offloading is affected by the doubly near-far problem, as expected, whereas the energy harvested can be used to fully operate local computation. Note that the expression for the success probability of offloading the entire task is given by setting $\mu = 1$, which results in $\phi_c = 0$. Similarly, for $\mu = 0$ and $t_e = 1$, we end up with the success probability of locally computing the entire task.

To maximize the success probability, the harvesting duration $t_e$ and the task split $\mu$ need to be optimized. In other words, we need to
\begin{align}
& \max_{t_e,\mu} ~ P_s(\mu,t_e), ~ \text{subject to} ~ 0 \leq t_e \leq 1, ~ 0 \leq \mu \leq 1,\label{muopt}
\end{align}
for given task of size $\ell$ bits. The optimal values of $\mu^*$ and $t_e^*$ can easily be derived numerically.

In order to simplify the expression in Theorem \ref{thm}, we provide a lower bound on the success probability. Based on the Taylor series expansion, we have that $\sqrt{1+x} \leq 1+x/2$. By applying this to \eqref{sp} and with the help of \cite[3.324-1]{GRAD}, we get
\begin{align}\label{lb}
P_s(\mu,t_e) \geq 2 \exp \left(-\frac{\phi_c r^\al}{t_e\gamma_2P} \right) \sqrt{\frac{\phi_or^\al}{t_e\gamma_2P}} K_1\left(2 \sqrt{\frac{\phi_or^\al}{t_e\gamma_2P}}\right),
\end{align}
where $K_1(\cdot)$ is the modified Bessel function of the second kind. Note that equality is achieved for small values of $\phi_c$ and $\phi_o$. Also, it is interesting to remark that the above bound is independent of the rectenna's parameter $\gamma_4$.

Next, in Proposition \ref{prop1} we derive the average number of successfully computed bits, given a certain allocation of the harvested energy to each process. Specifically, we assume that $\epsilon_o = \tau \epsilon$ is dedicated for offloading and the rest, i.e. $\epsilon_c = (1-\tau) \epsilon$, for local computation, $0 \leq \tau \leq 1$.

\begin{proposition}\label{prop1}
The average number of offloaded and locally computed bits, given that $\epsilon_o = \tau\epsilon$ and $\epsilon_c = (1-\tau)\epsilon$ is
\begin{align}
  \bar{\ell}(\tau,t_e) = \frac{t_d B}{\ln 2} &\int_0^\infty \exp\left(\frac{t_d r^{\al}\sigma^2}{\tau t_e h}\right) E_1\left(\frac{t_d r^{\al}\sigma^2}{\tau t_e h}\right) f_H(h) dh\nonumber\\
  &+ \left(\frac{(1-\tau) t_e}{\xi \psi^3}\right)^{\frac{1}{3}} \int_0^\infty h^{\frac{1}{3}} f_H(h) dh,
\end{align}
where $f_H(h)$ is given in Lemma \ref{lemma} with
$\theta_1 = \gamma_2 P r^{-\al}$ and $\theta_2 = \frac{3}{2} \gamma_4 P^2 r^{-2\al}$.
\end{proposition}

\begin{IEEEproof}
See Appendix \ref{prop1_prf}.
\end{IEEEproof}

From Proposition \ref{prop1}, we can obtain the expression for full offloading by setting $\tau = 1$ and for full local computation by setting $\tau = 0$. Similarly to above, in order to maximize the expected number of computed bits, the optimal values of $t_e^*$ and $\tau^*$ can be found by solving
\begin{align}
\max_{t_e,\tau} ~ \bar{\ell}(\tau,t_e), ~ \text{subject to} ~ 0 \leq t_e \leq 1, ~ 0 \leq \tau \leq 1.\label{tauopt}
\end{align}
As above, $t_e^*$ and $\tau^*$ can be evaluated numerically.

We now consider the case of deterministic fading. Based on \cite{BC}, non-deterministic fading can be beneficial to WPT. In what follows, we show that non-deterministic fading can benefit wireless powered offloading and local computation under specific scenarios. In the next proposition, we derive the maximum distance $r_{\text{max}}$, such that $\epsilon > \epsilon_o + \epsilon_c$, for the deterministic case.

\begin{proposition}\label{prop2}
The maximum distance $r_{\text{max}}$ up to which the device can offload $\ell_o$ bits and locally compute $\ell_c$ bits successfully, with deterministic fading, is
\begin{align}
r_{\text{max}} = \Big(&u + \left(v-\sqrt{v^2+(w-u^2)^3}\right)^\frac{1}{3}\nonumber\\
&\qquad\qquad+ \left(v+\sqrt{v^2+(w-u^2)^3}\right)^\frac{1}{3}\Big)^\frac{1}{\al},
\end{align}
where $u = -b/3a$, $v = u^3 + (bc - 3ad)/6a^2$, $w = c/3a$, $a \triangleq \left(2^\frac{\ell_o}{t_d B}-1\right) t_d \sigma^2,$ $b \triangleq \ell_c^3\xi\psi^3,$ $c \triangleq - t_e \gamma_2 P$ and $d \triangleq - \frac{3}{2} t_e \gamma_4 P^2.$
\end{proposition}

\begin{IEEEproof}
See Appendix \ref{prop2_prf}.
\end{IEEEproof}

Due to the binary nature of the deterministic case, $r \leq r_{\text{max}}$ implies that $P_s(\mu,t_e) = 1$, otherwise $P_s(\mu,t_e) = 0$. In the numerical results section below, we show that non-deterministic fading provides non-zero success probability for $r > r_{\text{max}}$. However, to further support our claim, we prove this analytically for the specific case of local computation. In this case, the maximum distance for the deterministic fading scenario is
\begin{align}
	r_{\rm max}^\alpha = \frac{P}{2\ell^3\xi\psi^3} \left(\gamma_2 + \sqrt{\gamma_2^2+6\gamma_4\ell_c^3\xi\psi^3}\right),
\end{align}
which follows from \eqref{cubic} with $a=0$ (solution to quadratic equation). From Theorem \ref{thm}, the success probability of local computation is
\begin{align}
P_s(0,1) = \exp\left(\frac{r^\alpha\gamma_2}{3\gamma_4 P}\left(1-\sqrt{1+\frac{6\gamma_4}{\gamma_2^2}\ell_c^3 \xi \psi^3}\right)\right).
\end{align}
Assume that we have a success probability equal to some $\delta \in [0,1]$. Then, by solving for $r^\alpha$, we get
\begin{align}
	r^\alpha = \frac{3\gamma_4 P \ln \delta }{\gamma_2} \left(1 - \sqrt{1+\frac{6\gamma_4}{\gamma_2^2}\ell_c^3\xi\psi^3}\right)^{-1}.
\end{align}
It suffices to show that there exists an $r$ such that $r^\alpha > r_{\rm max}^\alpha$. After simple algebraic operations, we have that
$r^\alpha > r_{\rm max}^\alpha \implies \delta < \exp(-1) = 0.3679$.
This means that, with non-deterministic fading, the device can locally compute its task at distances greater that $r_{\rm max}$, with a success probability less that $\exp(-1)$.

\section{Numerical Results}\label{nunerical}
In this section, we evaluate our considered system model and verify our mathematical analysis with computer simulations. We have used the following parameters, unless otherwise stated: $\al = 3$, $\gamma_2 = 0.0034$, $\gamma_4 = 0.3829$, $\xi = 10^{-28}$, $\psi = 10^3$ cycles/bit, $P = 0$ dB, and $\sigma^2 = -50$ dBm. In all figures, the mathematical analysis (lines) matches the simulation results (markers), which validates our theoretical approach.

\begin{figure}[t]\centering
  \includegraphics[width=0.9\linewidth]{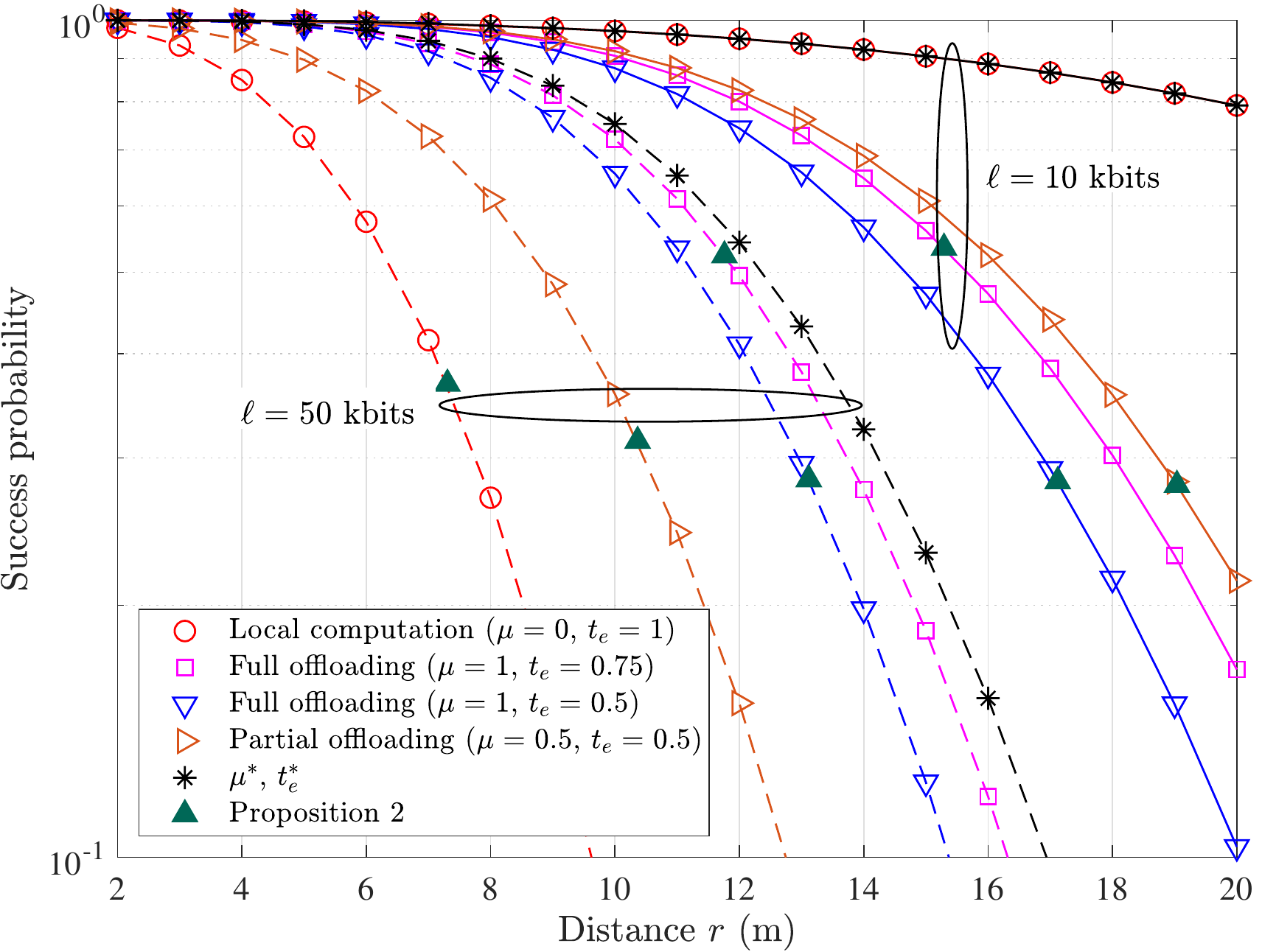}
  \caption{Success probability versus distance $r$; $B = 1$ MHz. Lines and markers depict theoretical and simulation results, respectively.}\label{fig1}
\end{figure}

\begin{figure}[t]\centering
	\includegraphics[width=0.9\linewidth]{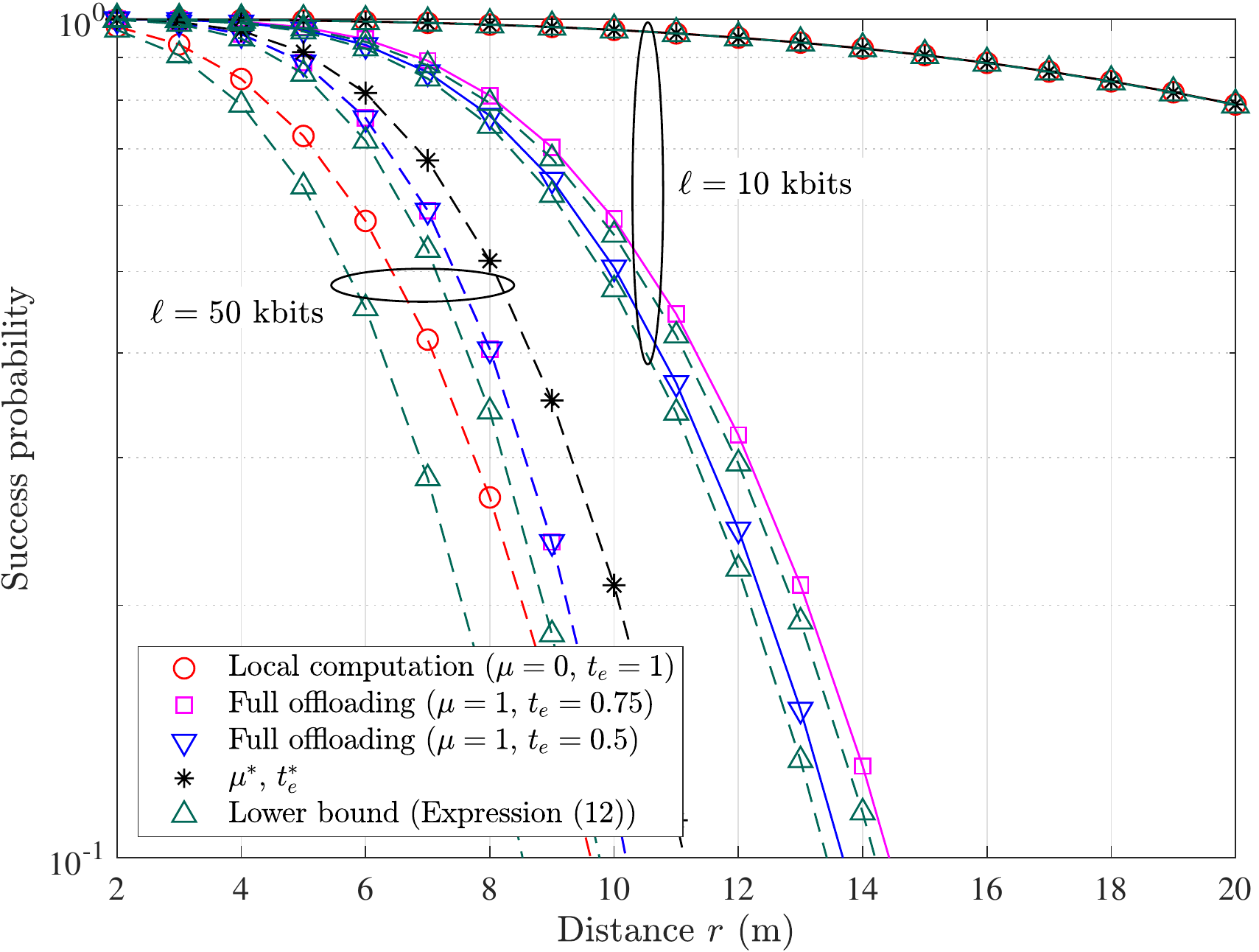}
	\caption{Success probability versus distance $r$; $B = 0.1$ MHz. Lines and markers depict theoretical and simulation results, respectively.}\label{fig2}
\end{figure}

Fig. \ref{fig1} depicts the success probability in terms of the distance $r$ for $B = 1$ MHz. As expected, the performance is critically affected by the distance $r$ to the AP, since it decreases the amount of harvested energy at the device. For $\ell = 10$ kbits, local computation outperforms both partial and full offloading schemes, especially for large values of $r$. Indeed, the success probability for full offloading drops faster due to the doubly near-far problem. Moreover, it is clear that the success probability with optimal $\mu^*$ and $t_e^*$ (based on \eqref{muopt}), corresponds to the local computation scheme; in other words, local computation is optimal for small tasks. On the other hand, for a larger task size ($\ell = 50$ kbits), local computation provides the worse performance and offloading is more preferable. Fig. \ref{fig1} also shows the maximum distance $r_{\text{max}}$ in a deterministic fading scenario (Proposition \ref{prop2}). The markers placed on the curves indicate that up to that point, all $\ell$ bits can be successfully computed. It is obvious that, when there is fading, successful computation can still be achieved even for distances larger that $r_{\text{max}}$, albeit with some probability. Note that, for the local computation case with $\ell = 50$ kbits, the marker is placed at $\exp(-1)$, which validates our analytical results. Fig. \ref{fig2} illustrates the same scenarios but with $B = 0.1$ MHz. Similar observations with Fig. \ref{fig1} hold. However, due to the fact that the bandwidth is smaller, the success probability of full and partial offloading is much lower. Finally, Fig. \ref{fig2} shows the lower bound expression given by \eqref{lb}; it is clear that the bound becomes tighter at smaller values of $r$ and $\ell$, which validates our statements.

Fig. \ref{fig3} shows the expected number of bits that can be successfully computed versus the distance $r$. For $B = 1$ MHz (top sub-figure), partial offloading overtakes full offloading at $10$ m ($t_e = 0.25$) and $17$ m ($t_e = 0.75$). As the device moves away from the AP, it needs more energy to offload the data, so it is preferable to locally compute a portion of the allocated task. For $B = 0.1$ MHz (bottom sub-figure), partial offloading outperforms both full offloading cases at shorter distances. As observed in Figs. \ref{fig1} and \ref{fig2}, less bandwidth implies that fewer number of bits can be offloaded and thus local computation starts to become more preferable. Finally, the optimal performance derived from $\eqref{tauopt}$, selects full or partial offloading for short distances and switches to local computation at larger distances.\vspace{-0.5mm}

\section{Conclusion}\label{conc}
In this paper, we investigated a point-to-point MEC system, where the device's offloading and/or local computation are powered solely by WPT. We considered a non-linear energy harvesting model and provided analytical closed-form expressions, which characterized the success probability and the average number of successfully computed bits. Our results showed that a combination scheme of partial offloading and local computation is not always efficient and the decision of whether to offload and/or compute locally, critically depends on the system's parameters.\vspace{-0.5mm}

\appendix
\subsection{Proof of Lemma \ref{lemma}}\label{lemma_prf}
The CDF $F_H(x)$ of $H$ is written as $F_H(x) = \PP(H < x) = \PP(\theta_1 |h|^2 + \theta_2 |h|^4 < x)$.
Then, by solving the quadratic inequality $\theta_1 |h|^2 + \theta_2 |h|^4 - x < 0$, we have
\begin{align}\label{sol}
|h|^2 < \frac{-\theta_1 \pm \sqrt{\theta_1^2+4\theta_2x}}{2\theta_2}.
\end{align}
Since $|h|^2$ is positive,
\begin{align}
\PP(\theta_1|h|^2 + \theta_2|h|^4 < x) = \PP\left(|h|^2 < \frac{-\theta_1 + \sqrt{\theta_1^2+4\theta_2x}}{2\theta_2}\right),
\end{align}
and the CDF follows from the fact that $|h|^2$ is an exponential random variable. Then, the PDF is derived by the derivative of the CDF. 

\subsection{Proof of Theorem \ref{thm}}\label{thm_prf}
By substituting \eqref{eh} in \eqref{ps}, we have
\begin{align}\label{p1}
P_s(\mu,t_e) = \PP\left(\gamma_2 |h|^2 \!+\! \frac{3}{2} \frac{\gamma_4 P}{r^\al} |h|^4 > \frac{r^\al}{t_e P}(\epsilon_o + \epsilon_c)\right),
\end{align}
where $\epsilon_o$ and $\epsilon_c$ are given by \eqref{eo} and \eqref{ec}, respectively. Then, using the CDF from Lemma \ref{lemma}, we have
\begin{align}
&P_s(\mu,t_e)\nonumber\\
&= \E_{|g|^2}\!\!\left[\exp\!\left(\frac{r^\al}{3\gamma_4 P}\left(\gamma_2\!-\!\sqrt{\gamma_2^2\!+\!\frac{6\gamma_4}{t_e}\left(\frac{\phi_o}{|g|^2}\!+\!\phi_c\right)}\right)\right)\!\right],
\end{align}
where $\phi_o$ and $\phi_c$ are defined in Theorem \ref{thm}. As $|g|^2$ is exponentially distributed, the theorem is proven by using the PDF $f_{|g|^2}(g)=\exp(-g)$ and by simplifying the expression with some algebraic manipulations.

\begin{figure}[t]\centering
	\includegraphics[width=0.9\linewidth]{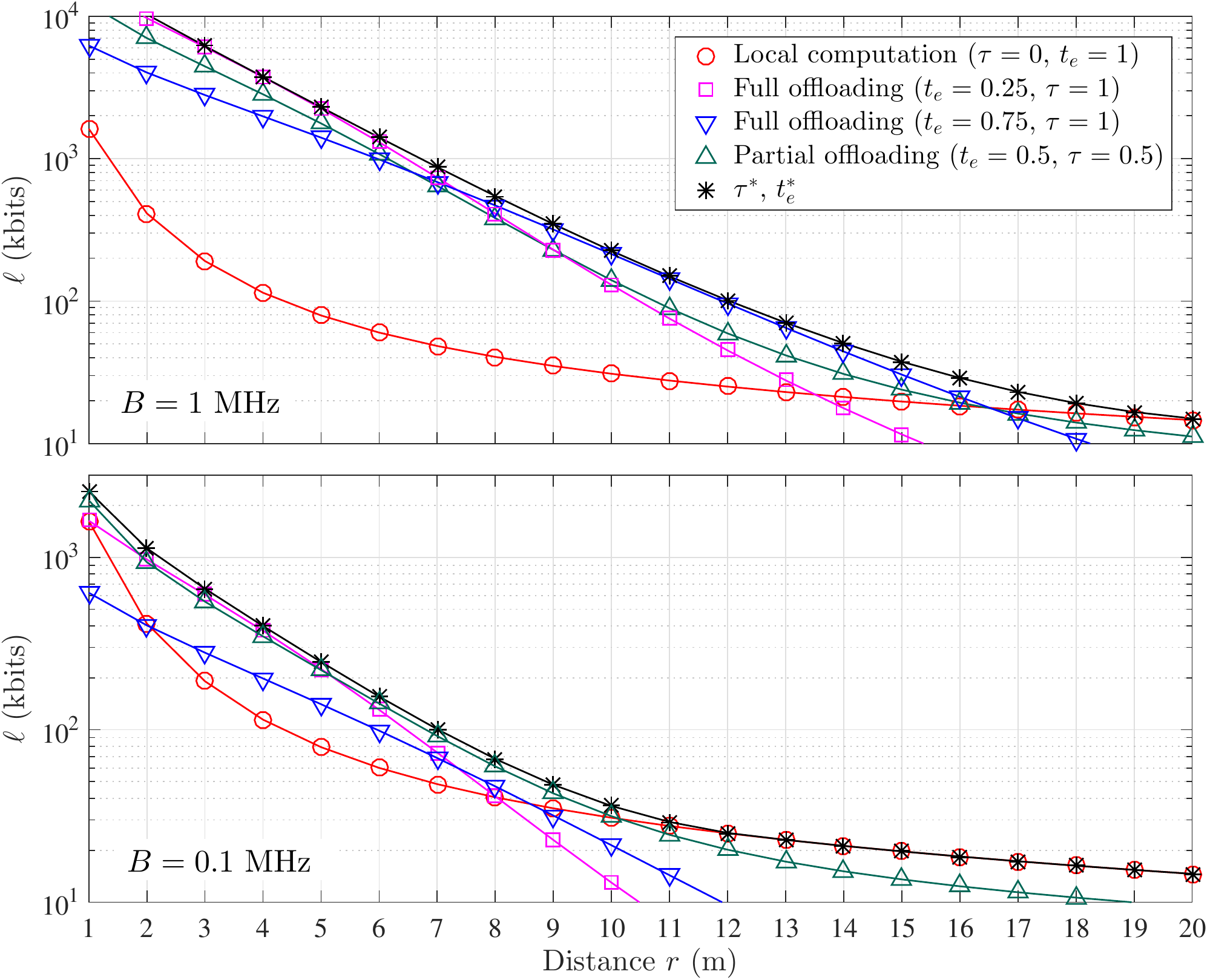}\vspace{-1mm}
	\caption{Expected number of computed bits $\bar{\ell}$ in terms of distance $r$; Lines and markers depict theoretical and simulation results, respectively.}\label{fig3}
\end{figure}

\subsection{Proof of Proposition \ref{prop1}}\label{prop1_prf}
The total number of bits $\ell$ offloaded and locally computed is $\ell = \ell_o + \ell_c$, with $\epsilon_o = \tau\epsilon$ and $\epsilon_c = (1-\tau)\epsilon$. The expectation of $\ell$ is then $\bar{\ell}(\tau,t_e) = \E[\ell] = \E[\ell_o] + \E[\ell_c]$. Hence,
\begin{align}
\E[\ell_o] &= t_d B \E_{|g|^2,\epsilon}\left[\log\left(1+\frac{\tau\epsilon}{t_d}\frac{|g|^2}{r^{\al}\sigma^2}\right)\right]\\
&= t_d B \E_{\epsilon}\left[\int_0^\infty \log\left(1+\frac{\tau\epsilon}{t_d}\frac{v}{r^{\al}\sigma^2}\right)\exp(-v) dv\right]\\
&= \frac{t_d B}{\ln 2} \E_{\epsilon}\left[\exp\left(\frac{t_d r^{\al}\sigma^2}{\tau\epsilon}\right) E_1\left(\frac{t_d r^{\al}\sigma^2}{\tau\epsilon}\right)\right],
\end{align}
which follows from $\log(x) = \ln(x)/\ln(2)$ and by employing \cite[4.337-2]{GRAD}; $E_1(x) = \int_x^\infty\exp(-t)t^{-1} dt$ is the exponential integral \cite{GRAD}. Thus, by replacing $\epsilon$ with \eqref{eh} and by using the PDF in Lemma \ref{lemma}, the result follows. Similarly,
\begin{align}
\E[\ell_c] &= \E\left[\left(\frac{\epsilon_c}{\xi \psi^3}\right)^{\frac{1}{3}}\right] = \left(\frac{1}{\xi \psi^3}\right)^{\frac{1}{3}} \E\left[\epsilon_c^{\frac{1}{3}}\right],
\end{align}
where $\epsilon_c = (1-\tau) \epsilon$ and $\epsilon$ is given by \eqref{eh}. By using the PDF in Lemma \ref{lemma}, we get the final expression.

\subsection{Proof of Proposition \ref{prop2}}\label{prop2_prf}
To successfully compute the specific task we need $\epsilon > \epsilon_o + \epsilon_c$. Therefore,
\begin{align}
&t_e \left(\frac{\gamma_2 P}{r^\al} + \frac{3}{2} \frac{\gamma_4 P^2}{r^{2\al}}\right) > \left(2^\frac{\ell_o}{t_d B}-1\right) t_d \sigma^2 r^\al + \ell_c^3\xi\psi^3\nonumber\\
&\implies \underbrace{\left(2^\frac{\ell_o}{t_d B}-1\right) t_d \sigma^2}_{\triangleq a} r^{3\al} + \underbrace{\ell_c^3\xi\psi^3}_{\triangleq b} r^{2\al} \underbrace{ - t_e \gamma_2 P}_{\triangleq c} r^\al\nonumber\\
&\hspace{5cm} \underbrace{- \frac{3}{2} t_e \gamma_4 P^2}_{\triangleq d} < 0,\label{cubic}
\end{align}
which is a cubic inequality with respect to $r^\al$. For the sake of simplicity, we define the coefficients as above. By using the substitution $r^\al = \rho - b/3a$, \eqref{cubic} is reduced to the depressed cubic\vspace{-1mm}
\begin{align}
\rho^3 + \rho\left(c-\frac{c^2}{3a}\right) + \frac{2b^3}{27a^2} - \frac{bc}{3a} + d < 0,
\end{align}
and thus using del Ferro's method \cite{GT}, the only real positive solution to the above cubic is
\begin{align}
\rho &< \left(v-\sqrt{v^2+(w-u^2)^3}\right)^\frac{1}{3} \!+\! \left(v+\sqrt{v^2+(w-u^2)^3}\right)^\frac{1}{3},
\end{align}
where $u = -b/3a$, $v = u^3 + (bc - 3ad)/6a^2$ and $w = c/3a$. By substituting back for $r$, the result follows.

\end{document}